\documentclass{appolb}
\usepackage{graphicx}
\usepackage{mathptmx}
\usepackage{geometry}
\usepackage{amsmath}
\usepackage{amssymb}
\usepackage{cite}
\newcommand{\ltapprox}{\raisebox{-0.5ex}{$\,\stackrel{<}{\scriptstyle\sim}\,$}}

\begin{document}

% ********************

\eqsec  % uncomment this line to get equations numbered by (sec.num)

\title{Simulations at fixed topology: fixed topology versus ordinary finite
volume corrections\thanks{Presented at Excited QCD 2015: Tatra Lominska, Slovakia}}

\author{Arthur Dromard$^{(1)}$, Wolfgang Bietenholz$^{(2)}$,\\ Urs Gerber$^{(2)}$, H\'{e}ctor Mej\'{\i}a-D\'{\i}az$^{(2)}$, Marc Wagner$^{(1)}$
\address{$^{(1)}$~Goethe-Universit\"at Frankfurt am Main \\
Institut f\"ur Theoretische Physik \\
Max-von-Laue-Stra{\ss}e 1, D-60438 Frankfurt am Main, Germany}
\address{$^{(2)}$~Instituto de Ciencias Nucleares \\
Universidad Nacional Aut\'{o}noma de M\'{e}xico \\
A.P.\ 70-543, C.P.\ 04510 Distrito Federal, Mexico}
}

\maketitle

\begin{abstract}
Lattice QCD simulations tend to get stuck in a single topological sector at fine lattice spacing, or when using chirally symmetric quarks. In such cases computed observables differ from their full QCD counterparts by finite volume corrections, which need to be understood on a quantitative level. We extend a known relation from the literature between hadron masses at fixed and at unfixed topology by incorporating in addition to topological finite volume effects, also ordinary finite volume effects. We present numerical results for SU(2) Yang-Mills theory.
\end{abstract}


% ********************

\section{Introduction}

In QCD simulations at small lattice spacings $a \ltapprox 0.05 \, \textrm{fm}$, algorithms typically have severe problems in generating transitions between different topological sectors. This problem of topology freezing is expected to appear for any lattice discretization of the quark and gluon fields \cite{Luscher:2011kk,Schaefer:2012tq}. For certain discretizations, e.g.\ chirally symmetric quarks, this problem is even present on coarser lattices\cite{Aoki:2008tq,Aoki:2012pma}. In specific cases it might be motivated to fix topology on purpose. For example, when using a mixed action setup with light overlap valence and Wilson sea quarks, one observes a rather ill-behaved continuum limit \cite{Cichy:2010ta,Cichy:2012vg}. This is due to near-zero modes of the Dirac operator in the valence sector, which are not compensated by corresponding modes in the sea. A possibility to circumvent this imbalance could be to restrict the lattice simulation to topological charge $Q=0$, where such near-zero modes are absent, e.g.\ by employing topology conserving actions \cite{Fukaya:2005cw,Bietenholz:2005rd,Bruckmann:2009cv}.

Methods to extract physically meaningful results from simulations at frozen or fixed topology have been proposed \cite{Brower:2003yx,Aoki:2007ka} and tested \cite{Bietenholz:2011ey,Bietenholz:2008rj,Bietenholz:2012sh,Dromard:2013wja,Czaban:2013haa,Bautista:2014tba,Dromard:2014ela,
Czaban:2014gva,Dromard:2014gma,Gerber:2014bia,Bautista:2015yza} in various models and theories. In this work we extend these methods by also including ordinary finite volume effects. Such a combined treatment of both fixed topology and ordinary finite volume corrections is expected to be particularly important for QCD at light $u/d$ quark masses. We test our equations in SU(2) Yang-Mills theory at fixed topology.

% ********************

\section{\label{SEC001}Topological finite volume effects}

In \cite{Brower:2003yx,Aoki:2007ka} an equation has been derived relating a hadron mass $M_{Q,V}$ obtained at fixed topological charge $Q$ and finite volume $V$ to its counterpart $M$ at unfixed topology (i.e.\ the physically meaningful hadron mass),
\begin{equation}
\label{EQN001} M_{Q,V} = M + \frac{1}{2 \chi_{t} V} M'' \bigg(1 -\frac{Q^2}{\chi_t V}\bigg) + \mathcal{O}\bigg(\frac{1}{(\chi_t V)^2}\bigg) ,
\end{equation}
where $M''$ denotes the second derivative of $M$ with respect to the $\theta$ angle at $\theta = 0$, and $\chi_t$ the topological susceptibility. This equation illustrates that fixed topology corrections are finite volume effects, i.e.\ effects suppressed by inverse powers of $V$. It is straightforward to extract a physical hadron mass $M$ from computations at fixed topology: one just has to fit eq.\ (\ref{EQN001}) to the available fixed topology and finite volume hadron masses $M_{Q,V}$, where $M$, $M''$ and $\chi_t$ are the fit parameters (examples of this procedure can be found in \cite{Bietenholz:2011ey,Bietenholz:2008rj,Bietenholz:2012sh,Dromard:2013wja,Czaban:2013haa,Bautista:2014tba,
Dromard:2014ela,Czaban:2014gva,Dromard:2014gma,Gerber:2014bia}).

\begin{figure}[htb]
\begin{centering}
\includegraphics[width=10.0cm]{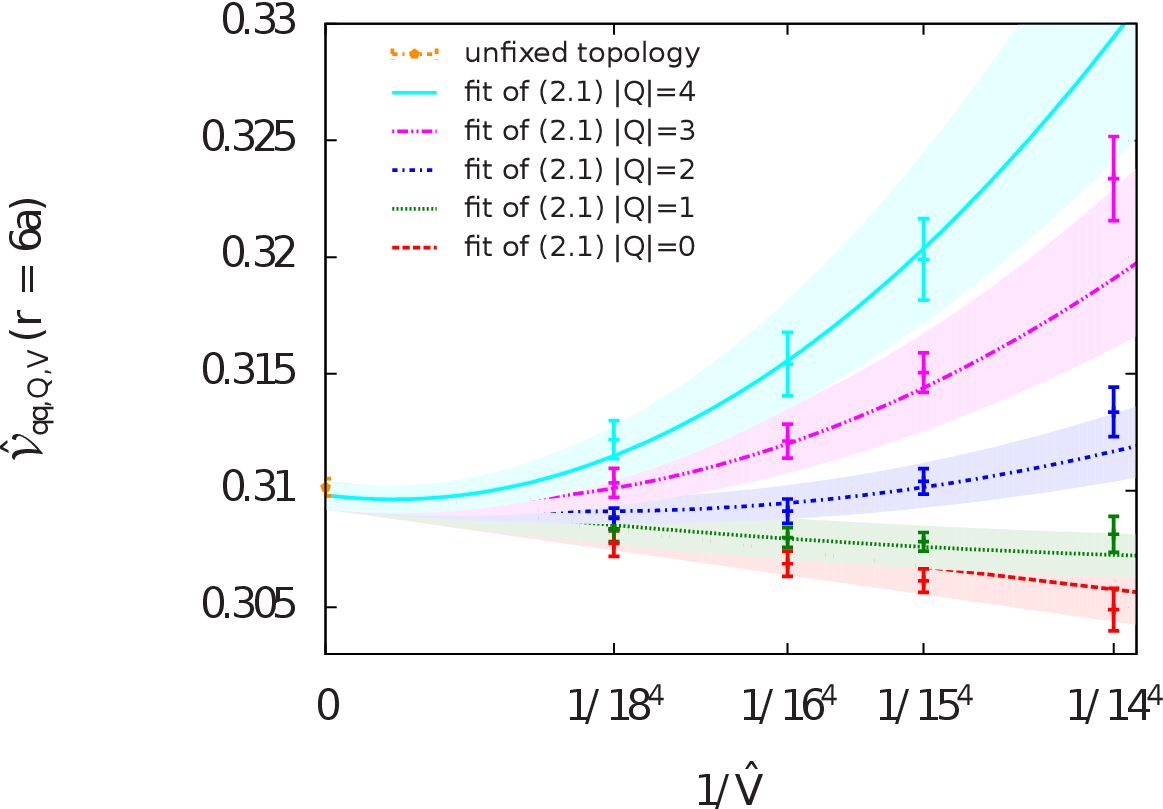}
\par
\end{centering}
\caption{\label{fig:result_TFV}$\hat{\mathcal{V}}_{q \bar{q},Q,V}(r = 6a)$ as a function of $1/\hat{V}$. The curves represent the fit of the lattice static potential results to eq.\ (\ref{EQN001}).}
\end{figure}

In Fig.~\ref{fig:result_TFV} we show recent results for SU(2) Yang-Mills theory (standard plaquette action, gauge coupling $\beta = 2.5$, i.e.\ lattice spacing $a \approx 0.073 \, \textrm{fm}$ \cite{Philipsen:2013ysa}). The static potential $\mathcal{V}_{q \bar{q},Q,V}$ at separation $r = 6a$ (which can be interpreted as a mass), has been computed in different topological sectors with topological charges $|Q| = 0,1,2,3,4$ and for different volumes $\hat{V} = 14^4, 15^4, 16^4, 18^4$ ($4000$ gauge link configurations have been generated for each of the four volumes). Note that the discrepancies between static potential results $\mathcal{V}_{q \bar{q},Q,V}(r = 6a)$ at different topological charges $Q$ are clearly visible, in particular for small volumes $V$ \footnote{Similar observations for the pion mass have been reported in \cite{Galletly:2006hq}.}. This demonstrates the necessity of using specific methods to properly deal with topological finite volume effects. The curves represent a single fit of eq.\ (\ref{EQN001}) to the lattice static potential results $\mathcal{V}_{q \bar{q},Q,V}(r = 6a)$. The fit is of good quality, i.e.\ $\chi^2 \ltapprox 1$. The resulting $\hat{\mathcal{V}}_{q \bar{q}}(r = 6a) = 0.3097(5)$ is in excellent agreement with a corresponding standard computation at unfixed topology, which yields $\hat{\mathcal{V}}_{q \bar{q}}(r = 6a) = 0.3101(3)$.

% ********************

\section{Ordinary finite volume effects}

Usually lattice simulations are performed at finite spatial volume $L^3$ with periodic boundary conditions. Consequently, a hadron at $\mathbf{x}$ will interact with images of itself, e.g.\ at $\mathbf{x} \pm L \mathbf{e}_x$, $\mathbf{x} \pm L \mathbf{e}_y$ or $\mathbf{x} \pm L \mathbf{e}_z$. Such interactions cause a shift in the hadron mass compared to infinite spatial volume, as first derived in \cite{Luscher:1985dn}.

\begin{figure}[htb]
\begin{centering}
\includegraphics[width=10.0cm]{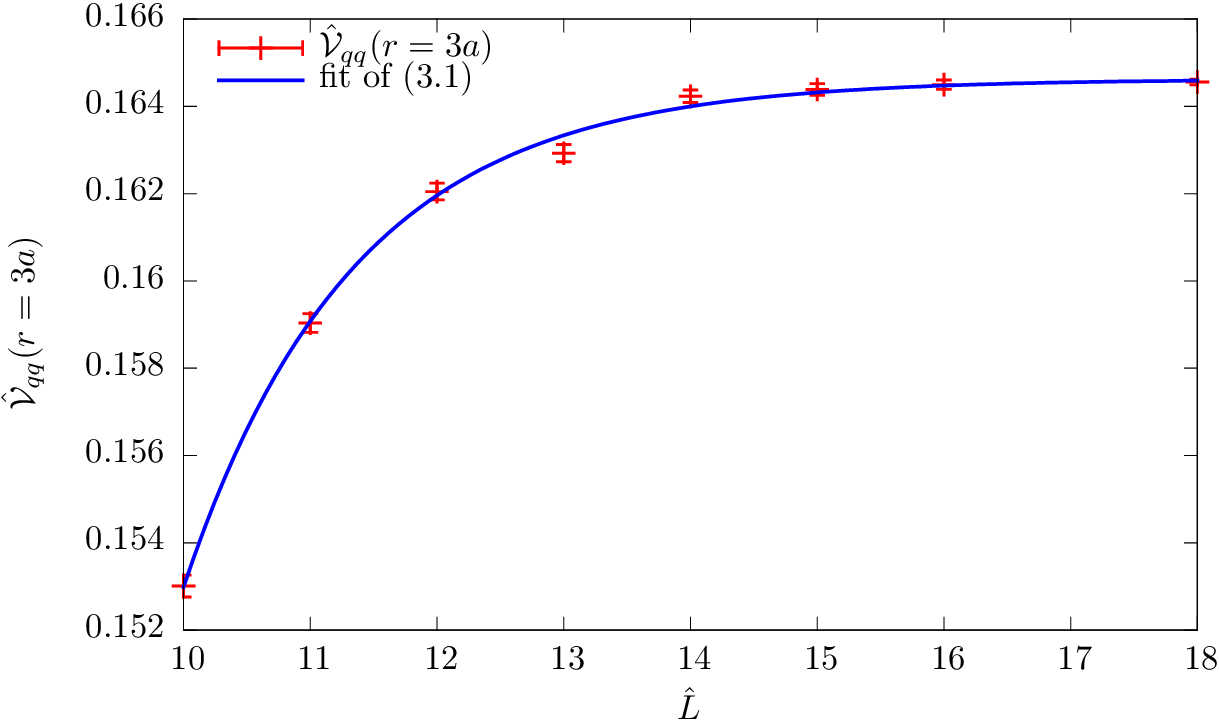}
\par
\end{centering}
\caption{\label{fig:OFT}The dependence of $\hat{\mathcal{V}}_{q \bar{q}}(r = 3a)$ on the periodic spatial extension $\hat{L}$ of the lattice (at unfixed topology).}
\end{figure}

The corresponding equation to describe these ordinary finite volume corrections (i.e.\ finite volume corrections not related to fixed topology) of the static potential in Yang-Mills theory is
\begin{equation}
\label{EQN002} M(L) - M(L \rightarrow \infty) \propto \frac{1}{L} \exp\bigg(-\frac{\sqrt{3} m L}{2}\bigg) ,
\end{equation}
where $M \equiv \mathcal{V}_{q \bar{q}}(r)$ and $m$ is the mass of the lightest particle, i.e.\ the $J^{PC} = 0^{++}$ glueball. In Fig.~\ref{fig:OFT} we confront this equation with lattice SU(2) Yang-Mills results for $\hat{\mathcal{V}}_{q \bar{q}}(r = 3a)$ and find excellent agreement (again we have used $\beta = 2.5$ and generated 4000 gauge link configurations for each of the eight volumes $\hat{V} = 10^4, 11^4, 12^4, 13^4, 14^4, 15^4, 16^4, 18^4$). For $\hat{L} \geq 14$ ordinary finite volume effects are negligible. For smaller $\hat{L}$, however, there are sizeable corrections, which need to be taken into account, in particular when using such volumes for computations at fixed topology as presented in the previous section. From the fit of eq.\ (\ref{EQN002}) to the lattice results shown in Fig.~\ref{fig:OFT}, one can even extract the $J^{PC} = 0^{++}$ glueball mass with remarkable precision, $\hat{m} = 0.74(4)$. This is in perfect agreement with the result obtained by a standard lattice computation of a glueball 2-point correlation function, $\hat{m} = 0.723(23)$ \cite{Teper:1998kw}.

% ********************

\section{Combining topological and ordinary finite volume effects}

In the SU(2) example discussed in Section~\ref{SEC001} it has been possible to analyze fixed topology results using eq.\ (\ref{EQN001}) in a meaningful way, i.e.\ without taking ordinary finite volume effects into account. Since the mass of the lightest particle, the $J^{PC} = 0^{++}$ glueball, is quite large, ordinary finite volume effects are strongly suppressed for large volumes. As indicated by Fig.~\ref{fig:OFT} and as done in Section~\ref{SEC001}, one just has to discard volumes with $\hat{L} < 14$. In QCD the situation is expected to be more difficult, because there the lightest particle, the pion, is much lighter than the $J^{PC} = 0^{++}$ glueball of SU(2) Yang-Mills theory. Moreover, lattice simulations of QCD, in particular at large volumes, are extremely demanding with respect to high performance computer resources. Therefore, it is highly desirable to combine eqs.\ (\ref{EQN001}) and (\ref{EQN002}), i.e.\ to obtain an expression describing both topological and ordinary finite volume corrections to hadron masses.

To derive such an expression, one has to consider ordinary finite volume effects also at non-vanishing $\theta$ angles using equations analogous to (\ref{EQN002}). These equations are the starting point for a lengthy calculation similar to that leading to eq.\ (\ref{EQN001}) (a detailed derivation of eq.\ (\ref{EQN001}) can e.g.\ found in \cite{Dromard:2014ela}). The resulting expression describing both topological and ordinary finite volume effects takes the form
\begin{equation}
\label{EQN003} \begin{aligned}
 & M_{Q,V} = M + \frac{1}{2 \chi_t V} M'' \bigg(1 - \frac{Q^2}{\chi_t V}\bigg) \\
 & \quad - \frac{A}{L} \bigg(1 + \frac{1}{2 \chi_t V} \bigg(\frac{A''}{A} - \sqrt{3} m'' L\bigg) \bigg(1 - \frac{Q^2}{\chi_t V}\bigg)\bigg) \exp\bigg(-\frac{\sqrt{3} m L}{2}\bigg) + \mathcal{O}\bigg(\frac{1}{(\chi_t V)^2}\bigg) ,
\end{aligned}
\end{equation}
where ordinary finite volume effects for the topological susceptibility $\chi_t$ have been neglected, since they are expected to be tiny \cite{DelDebbio:2004ns,Durr:2006ky}. Note that in addition to the parameters $M$, $M''$ and $\chi_t$, which are already present in eq.\ (\ref{EQN001}), there are four more parameters, $m$, $m''$, $A$ and $A''$, characterizing combined topological and ordinary finite volume corrections.

\begin{figure}[htb]
\begin{centering}
\includegraphics[width=9.5cm]{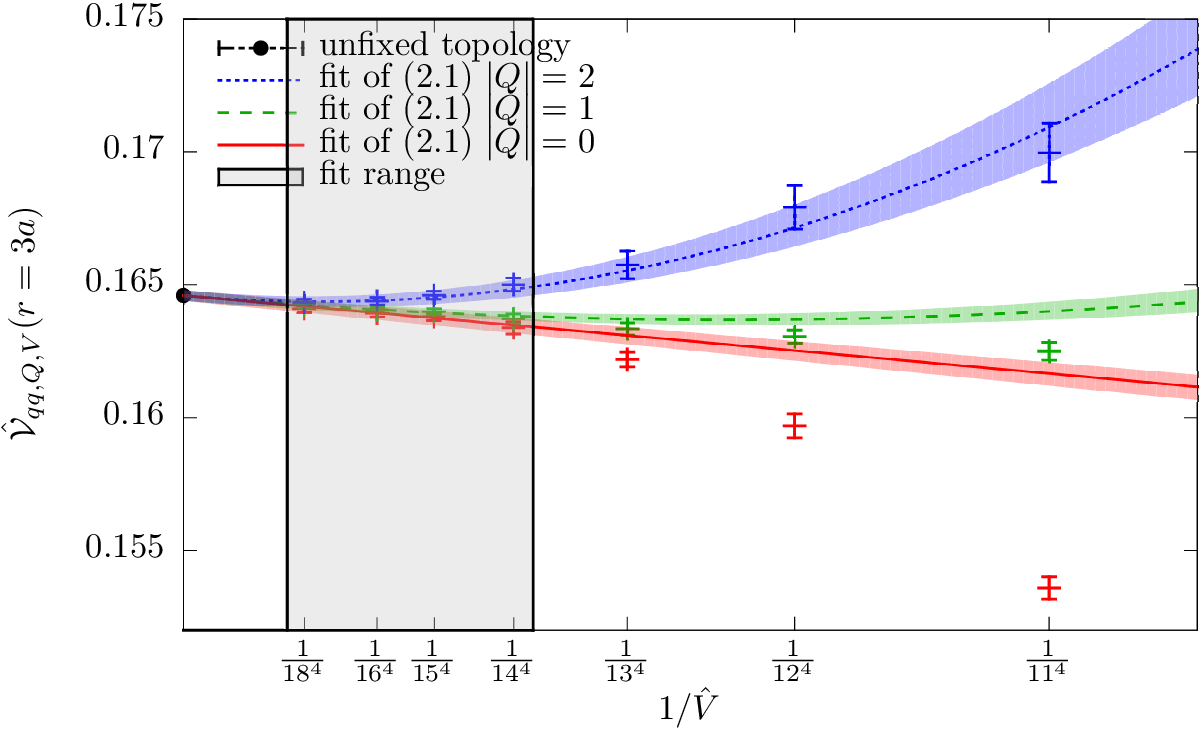}
\par
\includegraphics[width=9.5cm]{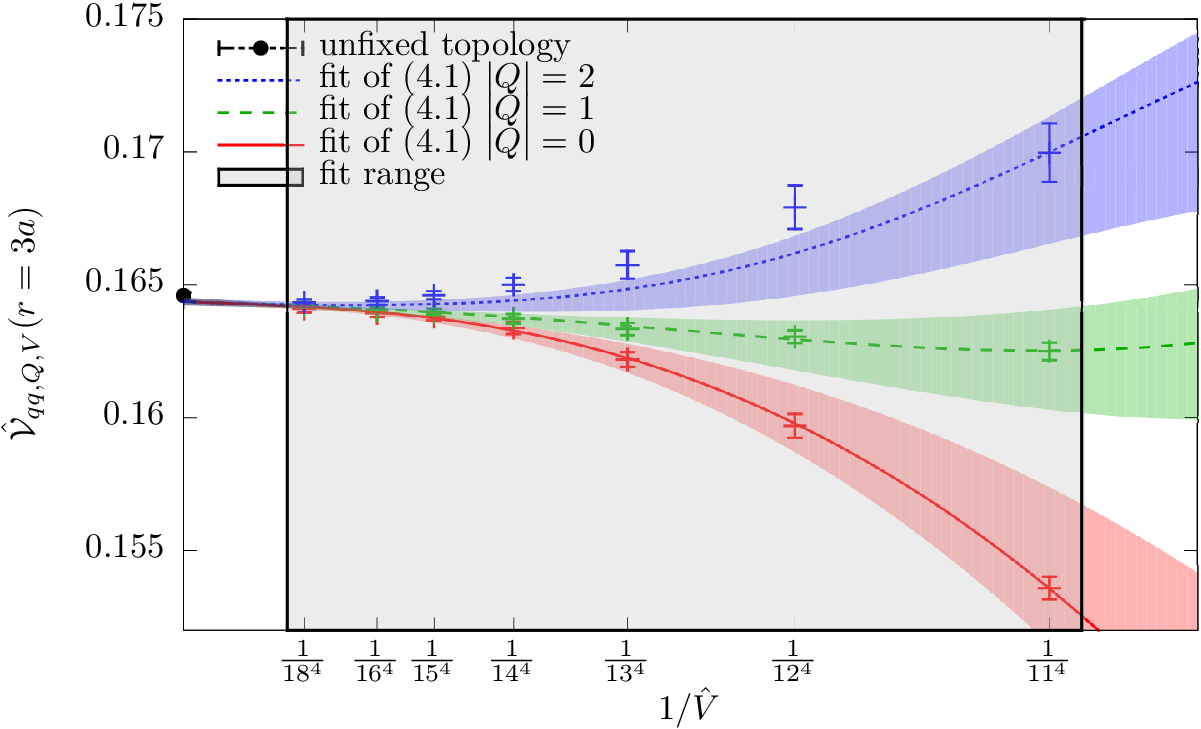}
\par
\end{centering}
\caption{\label{FIG003}$\hat{\mathcal{V}}_{q \bar{q},Q,V}(r = 3a)$ as a function of $1/\hat{V}$. \textbf{(top):}~The curves represent the fit of eq.\ (\ref{EQN001}) to the lattice static potential results for large volumes $\hat{V} = 14^4, 15^4, 16^4, 18^4$. There is a strong discrepancy between these curves and the lattice results for the small volumes $\hat{V} = 11^4, 12^4, 13^4$. \textbf{(bottom):}~The curves represent the fit of eq.\ (\ref{EQN003}) to the lattice static potential results for all volumes $\hat{V} = 11^4, \dots, 18^4$. There is almost perfect agreement, even at small volumes and $Q = 0$.}
\end{figure}

In Fig.~\ref{FIG003} (top) we show a plot similar to that from Fig.~\ref{fig:result_TFV}, this time for $\hat{\mathcal{V}}_{q \bar{q},Q,V}(r = 3a)$. Moreover, also results for small volumes $\hat{V} = 11^4, 12^4, 13^4$ are included. The curves correspond to eq.\ (\ref{EQN001}) with the fit parameters $M$, $M''$ and $\chi_t$ determined by a fit to the large volumes $\hat{V} = 14^4, 15^4, 16^4, 18^4$, where ordinary finite volume effects are negligible. There is a strong discrepancy between these curves and the lattice results for the small volumes $\hat{V} = 11^4, 12^4, 13^4$. This is expected, since ordinary finite volume corrections are not part of eq.\ (\ref{EQN001}) in particular for small Q.

\begin{table}[htb]
\begin{centering}
\begin{tabular}{|l|c|c|c|}
\hline 
 & $\hat{\mathcal{V}}_{q \bar{q},Q,V}(r = 3a)$ & $\hat{m}$ & $\hat{\chi_t} \times 10^5$ \tabularnewline
\hline 
fit results, eq.\ (\ref{EQN003}) & $0.16437(15)$ & $0.67(10)\phantom{0}$ & $9.5(2.0)$ \tabularnewline
unfixed topology results \cite{Teper:1998kw,deForcrand:1997sq} & $0.16455(7)\phantom{0}$ & $0.723(23)$ & $7.0(0.9)$ \tabularnewline
\hline 
\end{tabular}
\par
\end{centering}
\caption{\label{TAB001}Results for the static potential $\hat{\mathcal{V}}_{q \bar{q}}(r = 3a)$, the mass $\hat{m}$ of the $J^{PC} = 0^{++}$ glueball, and the topological susceptibility $\hat{\chi_t}$, obtained by a fit of eq.\ (\ref{EQN003}) to fixed topology lattice results $\hat{\mathcal{V}}_{q \bar{q},Q,V}(r = 3a)$.}
\end{table}

In Fig.~\ref{FIG003} (bottom) we show the same lattice results for $\hat{\mathcal{V}}_{q \bar{q},Q,V}(r = 3a)$. This time, however, the curves correspond to eq.\ (\ref{EQN003}) with the fit parameters $M$, $M''$, $\hat{\chi_t}$, $m$, $m''$, $A$ and $A''$ determined by a fit to all seven volumes $\hat{V} = 11^4, 12^4, 13^4, 14^4, 15^4, 16^4, 18^4$. There is almost perfect agreement, even at small volumes and for $Q = 0$. The extracted ``hadron mass'' $\hat{\mathcal{V}}_{q \bar{q},Q,V}(r = 3a)$ is consistent with a corresponding computation at unfixed topology and also the glueball mass $\hat{m}$ and the topological susceptibility $\hat{\chi_t}$ obtained by the fit are in fair agreement with reference values, cf.\ Table~\ref{TAB001}.

To conclude, we have incorporated ordinary finite volume corrections into an existing relation between hadron masses at fixed topology and physical hadron masses (i.e.\ hadron masses at unfixed topology). We have successfully tested the resulting equation in SU(2) Yang-Mills theory by studying the static potential at fixed topology. As an outlook, we plan to extend these tests to QCD in the near future.

% ********************

\section*{Acknowledgments}

  A.D. and M.W. acknowledge support by the Emmy Noether Programme of the DFG (German Research Foundation), grant WA 3000/1-1. 

 W.B., U.G. and H.M.-D. acknowledge support by the Consejo Nacional de Ciencia y Tecnolog\'{\i}a (CONACYT), project 155905/10, and DGAPA-UNAM, grant IN107915.
 
 This work was supported in part by the Helmholtz International Center for FAIR within the framework of the LOEWE program launched by the State of Hesse. Calculations on the LOEWE-CSC high-performance computer of Johann Wolfgang Goethe-University Frankfurt am Main were conducted for this research. We would like to thank HPC-Hessen, funded by the State Ministry of Higher Education, Research and the Arts, for programming advice.

% ********************

% ********************


\begin{thebibliography}{99}

\bibitem{Luscher:2011kk}
  M.~L\"uscher and S.~Schaefer,
  %``Lattice QCD without topology barriers,''
  JHEP {\bf 1107}, 036 (2011)
  [arXiv:1105.4749 [hep-lat]].
  %%CITATION = ARXIV:1105.4749;%%

\bibitem{Schaefer:2012tq}
  S.~Schaefer,
  %``Status and challenges of simulations with dynamical fermions,''
  PoS LATTICE {\bf 2012}, 001 (2012)
  [arXiv:1211.5069 [hep-lat]].
  %%CITATION = ARXIV:1211.5069;%%

\bibitem{Aoki:2008tq}
  S.~Aoki {\it et al.} [JLQCD Collaboration],
  %``Two-flavor QCD simulation with exact chiral symmetry,''
  Phys.\ Rev.\ D {\bf 78}, 014508 (2008)
  [arXiv:0803.3197 [hep-lat]].
  %%CITATION = ARXIV:0803.3197;%%

\bibitem{Aoki:2012pma}
  S.~Aoki {\it et al.},
  %``Simulation of quantum chromodynamics on the lattice with exactly chiral lattice fermions,''
  PTEP {\bf 2012}, 01A106 (2012).

\bibitem{Cichy:2010ta}
  K.~Cichy, G.~Herdoiza and K.~Jansen,
  %``Continuum Limit of Overlap Valence Quarks on a Twisted Mass Sea,''
  Nucl.\ Phys.\ B {\bf 847}, 179 (2011)
  [arXiv:1012.4412 [hep-lat]].
  %%CITATION = ARXIV:1012.4412;%%

\bibitem{Cichy:2012vg}
  K.~Cichy {\it et al.},
  %``Overlap valence quarks on a twisted mass sea: a case study for mixed action Lattice QCD,''
  Nucl.\ Phys.\ B {\bf 869}, 131 (2013)
  [arXiv:1211.1605 [hep-lat]].
  %%CITATION = ARXIV:1211.1605;%%

\bibitem{Fukaya:2005cw}
  H.~Fukaya {\it et al.},
  %``Topology conserving gauge action and the overlap-Dirac operator,''
  Phys.\ Rev.\ D {\bf 73}, 014503 (2006)
  [hep-lat/0510116].
  %%CITATION = HEP-LAT/0510116;%%

\bibitem{Bietenholz:2005rd}
  W.~Bietenholz {\it et al.},
  %``Exploring topology conserving gauge actions for lattice QCD,''
  JHEP {\bf 0603}, 017 (2006)
  [hep-lat/0511016].
  %%CITATION = HEP-LAT/0511016;%%

\bibitem{Bruckmann:2009cv} 
  F.~Bruckmann {\it et al.},
  %``Comparing topological charge definitions using topology fixing actions,''
  Eur.\ Phys.\ J.\ A {\bf 43}, 303 (2010)
  [arXiv:0905.2849 [hep-lat]].
  %%CITATION = ARXIV:0905.2849;%%

\bibitem{Brower:2003yx}
  R.~Brower, S.~Chandrasekharan, J.~W.~Negele and U.-J.~Wiese,
  %``QCD at fixed topology,''
  Phys.\ Lett.\ B {\bf 560}, 64 (2003)
  [hep-lat/0302005].
  %%CITATION = HEP-LAT/0302005;%%

\bibitem{Aoki:2007ka}
  S.~Aoki, H.~Fukaya, S.~Hashimoto and T.~Onogi,
  %``Finite volume QCD at fixed topological charge,''
  Phys.\ Rev.\ D {\bf 76}, 054508 (2007)
  [arXiv:0707.0396 [hep-lat]].
  %%CITATION = ARXIV:0707.0396;%%
  
%\cite{Bietenholz:2011ey}
\bibitem{Bietenholz:2011ey} 
  W.~Bietenholz, I.~Hip, S.~Shcheredin and J.~Volkholz,
  %``A Numerical Study of the 2-Flavour Schwinger Model with Dynamical Overlap Hypercube Fermions,''
  Eur.\ Phys.\ J.\ C {\bf 72}, 1938 (2012)
  [arXiv:1109.2649 [hep-lat]].
  %%CITATION = ARXIV:1109.2649;%%
  %16 citations counted in INSPIRE as of 24 Apr 2015

\bibitem{Bietenholz:2008rj}
  W.~Bietenholz and I.~Hip,
  %``Topological Summation of Observables Measured with Dynamical Overlap Fermions,''
  PoS LATTICE {\bf 2008}, 079 (2008)
  [arXiv:0808.3049 [hep-lat]].
  %%CITATION = ARXIV:0808.3049;%%

\bibitem{Bietenholz:2012sh}
  W.~Bietenholz and I.~Hip,
  %``Topological Summation in Lattice Gauge Theory,''
  J.\ Phys.\ Conf.\ Ser.\ {\bf 378}, 012041 (2012)
  [arXiv:1201.6335 [hep-lat]].
  %%CITATION = ARXIV:1201.6335;%%

\bibitem{Dromard:2013wja}
  A.~Dromard and M.~Wagner,
  %``Studying and removing effects of fixed topology in a quantum mechanical model,''
  PoS LATTICE {\bf 2013}, 339 (2014)
  [arXiv:1309.2483 [hep-lat]].
  %%CITATION = ARXIV:1309.2483;%%

\bibitem{Czaban:2013haa}
  C.~Czaban and M.~Wagner,
  %``Lattice study of the Schwinger model at fixed topology,''
  PoS LATTICE {\bf 2013}, 465 (2013)
  [arXiv:1310.5258 [hep-lat]].
  %%CITATION = ARXIV:1310.5258;%%
  
  %\cite{Bautista:2014tba}
\bibitem{Bautista:2014tba} 
  I.~Bautista {\it et al.},
  %``Interpretation of topologically restricted measurements in lattice sigma-models,''
  arXiv:1402.2668 [hep-lat].
  %%CITATION = ARXIV:1402.2668;%%
  %5 citations counted in INSPIRE as of 24 Apr 2015

\bibitem{Dromard:2014ela}
  A.~Dromard and M.~Wagner,
  %``Extracting hadron masses from fixed topology simulations,''
  Phys.\ Rev.\ D {\bf 90}, 074505 (2014)
  [arXiv:1404.0247 [hep-lat]].
  %%CITATION = ARXIV:1404.0247;%%

\bibitem{Czaban:2014gva}
  C.~Czaban, A.~Dromard and M.~Wagner,
  %``Studying and removing effects of fixed topology,''
  Acta Phys.\ Polon.\ Supp.\ {\bf 7}, 551 (2014)
  [arXiv:1404.3597 [hep-lat]].
  %%CITATION = ARXIV:1404.3597;%%

\bibitem{Dromard:2014gma}
  A.~Dromard, C.~Czaban and M.~Wagner,
  %``Hadron masses from fixed topology simulations: parity partners and SU(2) Yang-Mills results,''
  PoS LATTICE {\bf 2014}, 321 (2014)
  [arXiv:1410.4333 [hep-lat]].
  
  %\cite{Gerber:2014bia}
\bibitem{Gerber:2014bia} 
  U.~Gerber {\it et al.},
  %``Extracting Physics from Topologically Frozen Markov Chains,''
  PoS LATTICE {\bf 2014} (2014)
  [arXiv:1410.0426 [hep-lat]].
  %%CITATION = ARXIV:1410.0426;%%
  %2 citations counted in INSPIRE as of 24 Apr 2015

\bibitem{Bautista:2015yza}
  I.~Bautista {\it et al.},
  %``Measuring the Topological Susceptibility in a Fixed Sector: Results for Sigma Models,''
  arXiv:1503.06853 [hep-lat].

\bibitem{Philipsen:2013ysa}
  O.~Philipsen and M.~Wagner,
  %``On the definition and interpretation of a static quark anti-quark potential in the colour-adjoint channel,''
  Phys.\ Rev.\ D {\bf 89}, 014509 (2014)
  [arXiv:1305.5957 [hep-lat]].
  %%CITATION = ARXIV:1305.5957;%%

\bibitem{Galletly:2006hq} 
  D.~Galletly {\it et al.},
  %``Hadron spectrum, quark masses and decay constants from light overlap fermions on large lattices,''
  Phys.\ Rev.\ D {\bf 75}, 073015 (2007)
  [hep-lat/0607024].
  %%CITATION = HEP-LAT/0607024;%%

\bibitem{Luscher:1985dn}
  M.~L\"uscher,
  %``Volume Dependence of the Energy Spectrum in Massive Quantum Field Theories. 1. Stable Particle States,''
  Commun.\ Math.\ Phys.\ {\bf 104}, 177 (1986).
  %%CITATION = CMPHA,104,177;%%

\bibitem{Teper:1998kw} 
  M.~J.~Teper,
  %``Glueball masses and other physical properties of SU(N) gauge theories in D = (3+1): A Review of lattice results for theorists,''
  hep-th/9812187.
  %%CITATION = HEP-TH/9812187;%%

\bibitem{DelDebbio:2004ns}
  L.~Del Debbio, L.~Giusti and C.~Pica,
  %``Topological susceptibility in the SU(3) gauge theory,''
  Phys.\ Rev.\ Lett.\ {\bf 94}, 032003 (2005)
  [hep-th/0407052].
  %%CITATION = HEP-TH/0407052;%%

\bibitem{Durr:2006ky}
  S.~D\"{u}rr, Z.~Fodor, C.~Hoelbling and T.~Kurth,
  %``Precision study of the SU(3) topological susceptibility in the continuum,''
  JHEP {\bf 0704}, 055 (2007)
  [hep-lat/0612021].
  %%CITATION = HEP-LAT/0612021;%%

\bibitem{deForcrand:1997sq}
  P.~de Forcrand, M.~Garc\'{i}a P\'{e}rez and I.~O.~Stamatescu,
  %``Topology of the SU(2) vacuum: A Lattice study using improved cooling,''
  Nucl.\ Phys.\ B {\bf 499}, 409 (1997)
  [hep-lat/9701012].
  %%CITATION = HEP-LAT/9701012;%%

\end{thebibliography}
\end{document}